\documentclass[12pt,reqno]{article}

\usepackage{amsfonts}
\usepackage{amsmath}
\usepackage{amssymb}
\usepackage{amsthm}
\usepackage{setspace}
\usepackage{fullpage}
\usepackage{graphicx}
\usepackage{url}
\usepackage{braket}
\usepackage{mathtools}
\usepackage[all]{xy}
\usepackage{caption}

\usepackage{color}
\usepackage{bm}

\usepackage[numbers]{natbib}

\usepackage{verbatim}
\usepackage{graphicx}

\usepackage{mathrsfs}

\newtheorem*{buhyp}{Baby Universe Hypothesis}

\theoremstyle{remark}

\theoremstyle{definition}

\newcommand{\Z}{\mathcal{Z}}

\newcommand{\CC}{\mathbb{C}}

\newcommand{\D}{\mathcal{D}}
\renewcommand{\L}{\mathcal{L}}

\newcommand{\del}{\partial}

\newcommand{\tensor}{\otimes}
\newcommand{\abs}[1]{\left\lvert #1 \right\rvert}

\renewcommand{\H}{\mathcal{H}}

\newcommand{\BU}{\text{BU}}
\newcommand{\HH}{\text{HH}}
\newcommand{\QG}{\text{QG}}
\newcommand{\CFT}{\text{CFT}}
\newcommand{\EFT}{\text{EFT}}

\begin{document}

\begin{titlepage}
\hfill \\
\vspace*{15mm}
\begin{center}
{\Large \bf Baby Universes, Holography, and the Swampland}

\vspace*{15mm}

{\large Jacob McNamara and Cumrun Vafa}
\vspace*{8mm}

Jefferson Physical Laboratory, Harvard University, Cambridge, MA 02138, USA\\

\vspace*{0.7cm}

\end{center}
\begin{abstract}

On the basis of a number of Swampland conditions, we argue that the Hilbert space of baby universe states must be one-dimensional in a consistent theory of quantum gravity. This scenario may be interpreted as a type of ``Gauss's law for entropy'' in quantum gravity, and provides a clean synthesis of the tension between Euclidean wormholes and a standard interpretation of the holographic dictionary, with no need for an ensemble. Our perspective relies crucially on the recently-proposed potential for quantum-mechanical gauge redundancies between states of the universe with different topologies. By an application of the state-operator correspondence, this proposal rules out the possibility of nontrivial, strictly well-defined bulk operators supported in a compact region. We further comment on the possible exceptions in $d\leq 3$ for this hypothesis, and the role of an ensemble for holographic theories in low dimensions, such as JT gravity in $d = 2$ and possible cousins in $d=3$. We argue that these examples are incomplete physical theories that should be viewed as branes in a higher dimensional theory of quantum gravity, for which an ensemble plays no role.

\end{abstract}

\end{titlepage}

\tableofcontents

\section{Introduction}

One of the basic lessons of string theory is that, in quantum gravity, anything that can be dynamical must be dynamical. In particular, every coupling constant in string theory is the asymptotic value of a dynamical field, and every symmetry is coupled to a dynamical gauge field. This is a version of background independence, since otherwise we would have a theory of quantum gravity that depends on a choice of fixed background parameters. The condition that quantum gravity must have no free parameters has been codified as one of the most basic swampland conditions (for a review of the Swampland Program, see \cite{BrennanCartaVafa,Palti:2019pca}), and has been proposed to hold in all consistent theories of quantum gravity in $d \geq 3 $ spacetime dimensions

The restriction to $d \geq 3$ dimensions seems essential, since the nature of gravity dramatically changes for $d\leq 3$.  Indeed there are obvious counterexamples in $d = 2$ to the lack of free parameters. In particular, any coupling constant of the worldsheet CFT (such as the radii of internal geometries) is a free parameter of the two-dimensional theory of quantum gravity on the string worldsheet. Related to this fact, note that in $d = 2$, massless scalar fields do not get a vev, and do not lead to different superselection sectors in infinite spatial volume.  Thus, we cannot hope to realize a parameter of the theory as the vacuum expectation value of a massless dynamical field.\footnote{Of course, a massive parameter could be realized as the vev of a massive field.} More generally, many swampland conditions are violated in $d = 2$, such as the absence of global symmetries and the triviality of cobordism. The case $d=3$ may admit similar exceptions because gravity has no propagating degrees of freedom, and there might exist topological theories of gravity (such as gravitational Chern-Simons theory) which do not follow the usual swampland conditions of higher dimensional quantum theories of gravity. To cover these potential exceptions in $d = 3$, we can replace the restriction of $d\geq 3$ with $d>3$. 

The principle that there are no free parameters in a quantum theory of gravity has come under question in recent studies of $d = 2$ holography in the context of JT gravity \cite{Saad:2019lba}. In particular, if one considers a collection of one-dimensional quantum systems, depending on some free parameters averaged over an ensemble, one obtains a holographic quantum gravity dual in $d = 2$, which entails summing over bulk geometries possibly connecting the different boundaries. Moreover, changing the parameters for this ensemble will change the parameters of the gravitational theory, and so the $d = 2$ bulk gravity would inherit free unfixed parameters. It is tempting to ask whether this lesson learned in $d = 2$ can be exported to higher dimensions, and should lead to a departure from the standard AdS/CFT dictionary in higher dimensions, forcing us to contend with an ensemble of dual boundary theories.

This ensemble average may be interpreted as the result of an old, well known source of free parameters in quantum gravity \cite{Saad:2019lba}, namely the so-called $\alpha$-parameters of Coleman \cite{Coleman}, studied further in \cite{GiddingsStrominger1, GiddingsStrominger2} (for a review and other connections to the Swampland Program, see \cite{HebeckerMikhailSoler}). In the Euclidean path integral, spacetime wormholes can be interpreted as calculating amplitudes to produce or absorb baby universes. These processes pose a threat to unitarity of the quantum system, in the form of potential information loss \cite{Coleman, GiddingsStrominger1} or non-factorization of correlation functions in a holographic dual \cite{{Arkani-HamedOrgeraPolchinski}}. The proposed resolution is to suppose the baby universes are in a specific $\alpha$-eigenstate (in which case there are no issues with unitarity and factorization) at the cost of introducing $\alpha$ as free parameters of the theory, which are not the expectation value of any dynamical fields. Thus, we see an immediate tension between the Euclidean path integral and the expectation from the Swampland Program that quantum gravity should have no free parameters.

Recently, Marolf and Maxfield have provided a beautiful analysis of $\alpha$-parameters and the baby universe Hilbert space in a two-dimensional toy model \cite{MarolfMaxfield}. While they do still find nontrivial $\alpha$-parameters, a key ingredient in their story is that the Euclidean path integral implies an enormous redundancy in the naive baby universe Hilbert space, leading to an enormous reduction in the number of degrees of freedom. Essentially, cobordisms connecting baby universe states with different topologies act as a generalized form of gauge transformation, which force the gauge-invariant wavefunctions to include specific superpositions of baby universes with different topologies, an idea which has also been considered in \cite{Jafferis}. Chief among the gauge invariant wavefunctions is the Hartle-Hawking wavefunction \cite{HartleHawking}, defined by the Euclidean path integral with no initial boundary, but there are many other gauge-invariant states in their toy model, leading to many different $\alpha$-eigenstates.

What, then, is the lesson for quantum gravity in dimension $d >3$? There seems to be a paradox: the general considerations leading to $\alpha$-parameters make no reference to the dimension, but the Swampland Program heavily suggests that there should be no free parameters in a theory of quantum theory of Einstein gravity in $d > 3$. However, there is a clean resolution, which is to suppose that the gauge redundancies described in \cite{MarolfMaxfield, Jafferis} are so strong in $d > 3$ that they collapse the entire baby universe Hilbert space to a single quantum state. If this were the case, there would be no nontrivial $\alpha$-parameters, and so there would be no tension between the Euclidean path integral and our understanding of quantum gravity in $d > 3$. In addition, as pointed out in \cite{MarolfMaxfield}, in this case there would be no need to discuss modifications of the standard holographic dictionary involving ensembles.

In this note, we argue that this indeed must happen in a consistent theory of quantum gravity in $d > 3$. That is, we propose the following hypothesis as a swampland condition.

\begin{buhyp}
Let $\H_\BU$ be the Hilbert space of baby universes in a unitary theory of quantum gravity in $d > 3$ spacetime dimensions. Then we have $\dim \H_\BU = 1$.
\end{buhyp}

\noindent Put differently, we propose that the \emph{only} gauge invariant state of baby universes is the Hartle-Hawking wavefunction. As a swampland condition, this hypothesis should place enormous constraints on which effective field theories admit a realization in quantum gravity. Indeed, the Baby Universe Hypothesis will be massively violated with a generic choice of matter and interactions, and looks quite miraculous from the perspective of effective field theory.

As a consequence, we conclude that the ensemble interpretation of holography is very much a feature of $d = 2$ and potentially $d=3$ theories of gravity, for which many swampland principles do not apply.   
Further, we explain why the presence of an ensemble in $d=2$ (including the case of JT gravity) is actually already {\it anticipated} by the standard (non-ensemble) holography of higher dimensional theories, by viewing a $d = 2$ spacetime as the `t Hooft worldsheet associated to the standard holographic duality of a higher dimensional gauge theory.\footnote{A similar perspective on the presence of an ensemble and its relationship with the computation of Wilson loop observables has appeared in \cite{BetziosPapadoulaki} in the context of $c = 1$ Liouville theory.} That there is no known analog of higher dimensional ($d>3$) branes that can serve as a large $N$ perturbative expansion of some QFT system is beautifully compatible with our hypothesis that we do not expect an ensemble interpretation of holography to exist for $d>3$.  
We argue that any such low dimensional exceptions, including JT gravity, are incomplete physical theories:  they should not be viewed as standalone quantum gravitational systems, but rather as worldvolume theories of branes in higher dimensional theories of quantum gravity that do not enjoy any such exceptions.\footnote{From this perspective, we believe that there is no exception to our hypothesis for a complete quantum gravitational system in any dimension. In particular, we expect that all quantum theories of gravity which arise by compactification of string theory down to $d=2$ and $3$ dimensions pose no exceptions to the condition of no free parameters.} In particular, the large ambiguity in choosing an ensemble is related to the ambiguity in the choice of which observable to measure in the higher-dimensional quantum system.

This note is organized as follows. In Section \ref{baby_universes}, we review Coleman's argument \cite{Coleman} for $\alpha$-parameters and their implications for unitarity, as well as the potential for an enormous gauge redundancy \cite{MarolfMaxfield, Jafferis} in the baby universe Hilbert space. In Section \ref{swampland}, we argue for the Baby Universe Hypothesis on the basis of a number of swampland conditions, as well as derive the absence of global symmetries as a consequence of the Baby Universe Hypothesis. Further, we use a state-operator correspondence to rephrase the Baby Universe Hypothesis as the absence of local bulk operators. In Section \ref{holography}, we describe the implications of the Baby Universe Hypothesis for holography, both in the context of AdS/CFT and more broadly as an interpretation of the general holographic principle as ``Gauss's law for entropy." We also discuss why the ensemble interpretation for holography in the case of $d=2$ (and potentially $3$) is natural, and why we do not expect this exception to persist in higher dimensions. Finally, in Section \ref{conclusion}, we conclude our discussion.

\section{Baby Universes and $\alpha$-Parameters}\label{baby_universes}

In this section, we review the arguments \cite{Coleman, GiddingsStrominger1, GiddingsStrominger2} that Euclidean wormholes and the production of baby universes can lead to an ensemble of quantum systems, labeled by so-called $\alpha$-parameters, as well as the potential loss and restoration of unitarity. Further, we review the picture outlined in \cite{MarolfMaxfield, Jafferis} of a potentially enormous gauge redundancy in quantum gravity relating states with different topologies, and its role in defining the baby universe Hilbert space in particular.

\subsection{$\alpha$-Parameters and Coupling Constants}\label{coleman}

In \cite{Coleman}, Coleman examines the fact that the Euclidean path integral naturally includes configurations that can be interpreted as tunneling amplitudes for a small piece of the universe to detach as a disconnected baby universe, only to reattach itself at another point in spacetime. Coleman outlines how attempting to integrate out the effects of these Euclidean wormholes leads to an ensemble of bulk theories, labeled by different $\alpha$-parameters, which label states of the baby universes in which no quantum coherence is lost. These $\alpha$-parameters correspond to coupling constants in the action of our effective field theory. In this section, we review Coleman's argument, highlighting the features that will play a key role for us below.

\vspace{12pt}
\begin{figure}[!ht]
\centering
\includegraphics[height = 2 in]{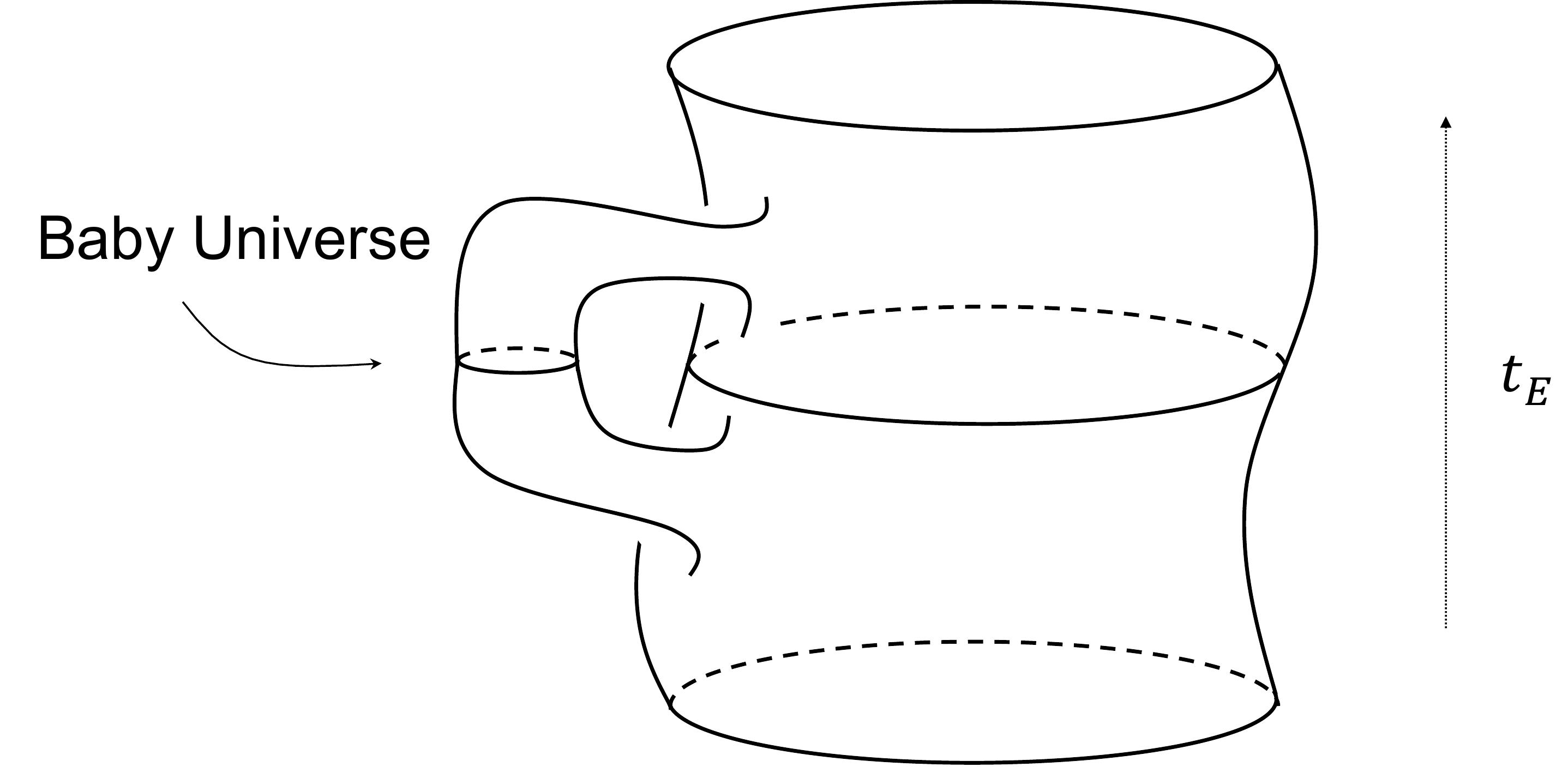}
\caption*{A baby universe detaches and reattaches in Euclidean time.}
\end{figure}

The basic argument presented in \cite{Coleman} is as follows.\footnote{For mathematicians, this argument may be summarized by saying that the baby universe Hilbert space is naturally a commutative Frobenius algebra under disjoint union, together with the structure theorem for such algebras.} Suppose that there are many species of baby universe, labeled by a discrete variable $i$. We may define baby universe creation and annihilation operators $a_i^\dagger, a_i$, which satisfy the standard bosonic commutation relations. In terms of these operators, the effective action may be written as
\begin{equation} S = S_0 + \sum_i (a_{i^*}^\dagger + a_i) \int \L_i, \end{equation}
where $i^*$ labels the CPT conjugate of the baby universe state $i$, and where the absorption of baby universe $i$ inserts the local operator $\L_i$ in the effective field theory. Notably, we have pulled the operators $a_{i^*}^\dagger, a_i$ outside of the integral over spacetime, as they are independent of position in spacetime (put differently, the baby universes carry zero momentum).

Now, we define
\begin{equation} A_i = a_{i^*}^\dagger + a_i. \end{equation}
The operators $A_i$ all mutually commute, which allows us to define mutual eigenstates $\ket{\alpha}$, where we have
\begin{equation} A_i \ket{\alpha} = \alpha_i \ket{\alpha}. \end{equation}
In addition, the operators $A_i$ all commute with the Hamiltonian, and so the different $\alpha$-eigenstates define different superselection sectors of the effective field theory. In a fixed $\alpha$-eigenstate, the effective action takes the form
\begin{equation} S = S_0 + \sum_i \alpha_i \int \L_i, \end{equation}
and so we see that the $\alpha$-parameters serve as coupling constants for different terms in the effective action.

\subsection{Potential Loss of Unitarity}

Suppose we begin in a state of the universe which is a tensor product of some fixed baby universe state and a state of the effective field theory, that is, suppose
\begin{equation} \ket{\psi} = \ket{\psi_\BU} \tensor \ket{\psi_\EFT}. \end{equation}
Generically, under time evolution, such a state will evolve into an entangled state,
\begin{equation} \ket{\psi'} = \sum_i c_i \ket{\psi_\BU^i} \tensor \ket{\psi_\EFT^i}. \end{equation}
If we trace out the baby universe state, we see that from the perspective of effective field theory, a pure state has evolved into a mixed state
\begin{equation} \ket{\psi_\EFT} \bra{\psi_\EFT} \rightsquigarrow \sum_i \abs{c_i}^2 \ket{\psi_\EFT^i} \bra{\psi_\EFT^i}, \end{equation}
and so information has been lost. Put differently, the state of the universe is becoming entangled with the state of the baby universes, which looks to an observer in the effective field theory like a loss of information \cite{AkersEngelhardtHarlow}.

Though in a strict sense information is being lost in this process, it has been argued that the existence of $\alpha$-eigenstates ensures that nothing catastrophic arises from this information loss. First of all, if
\begin{equation} \ket{\psi_\BU} = \ket{\alpha}, \end{equation}
is initially an $\alpha$-eigenstate, then since the Hamiltonian only interacts with the baby universe Hilbert space through the operators $A_i$, the baby universes will remain in an $\alpha$-eigenstate for all time, and so there will be no transfer of information to the baby universes. Thus, each different choice of $\alpha$-parameters defines a well-defined and unitary effective field theory, which may be thought of as different superselection sectors. These different effective field theories differ only in the values of their coupling constants.

What are we supposed to think if the baby universes are in a superposition of the different $\alpha$-eigenstates? In this case, information is indeed lost. However, though we lose track of the relative phases of the different $\alpha$-eigenstates encoded in the initial state $\ket{\psi_\BU}$, we could not detect this information to begin with. Since we can only interact with the baby universe Hilbert space through the mutually commuting operators $A_i$, we should think of a state $\ket{\psi_\BU}$ as only encoding a classical mixture of states with different $\alpha$. Put differently, the system describes a classical ensemble of different effective field theories, with a probability distribution on $\alpha$ corresponding to our ignorance of the exact values of coupling constants. Thus, while the results of quantum gravitational experiments (such as black hole formation and decay) cannot be predicted exactly, this will be interpreted as a lack of precise knowledge of coupling constants, as opposed to a contradiction with unitary time evolution.

Though this argument holds up to scrutiny, already we see that there is an another, much cleaner scenario for preserving unitarity. This is to suppose that for some mysterious reason, the Hilbert space of baby universes must be one-dimensional! In this case, the baby universes would have no capacity for storing information, and so even though information tries to lose itself in the state of the baby universes, the vanishing entropy of closed universes forces information to be preserved in the effective field theory! At this stage, this hypothesis merely trades discomfort at a formal, yet undetectable, loss of information in a theory with $\alpha$-parameters for a perhaps much greater discomfort at the idea that the baby universe Hilbert space is somehow one-dimensional. However, as we argue below, this is the natural conclusion from the perspective of the Swampland Program, and thus our discomfort can be interpreted as a relic of trying to apply the generic expectations of effective field theory to the non-generic case of quantum gravity!

\subsection{Gauge Redundancy in the Baby Universe Hilbert Space}\label{gauge}

How could it be that the Hilbert space $\H_\BU$ of baby universe states is one-dimensional? A naive count of degrees of freedom would suggest a much larger dimension, since from the perspective of effective field theory, there are fluctuating degrees of freedom attached to each point in space, and moreover there seem to be degrees of freedom corresponding to the topology of the baby universes. This is the basic tension between effective field theory and the holographic principle, in that there must be some mechanism in non-perturbative quantum gravity that greatly reduces the number of physically distinct configurations, in order to satisfy an area law for the entropy.

Such a mechanism has recently been proposed and studied in \cite{MarolfMaxfield, Jafferis}, which is to realize that there are gauge redundancies which can identify wavefunctions of the universe with different topologies. One place this can be seen clearly is in the case of the two sided AdS-Schwarzschild black hole, which is dual to the thermofield double state in AdS/CFT \cite{Maldacena}, a quantum superposition of states with disconnected spatial topology. As pointed out in \cite{Jafferis}, the fact that the two-sided black hole is not orthogonal to disconnected states may be seen from the Euclidean path integral, which includes configurations that change the topology of the spatial slice, leading to a nonzero inner product between states with different topology.

\vspace{12pt}
\begin{figure}[!ht]
\centering
\includegraphics[height = 2.2 in]{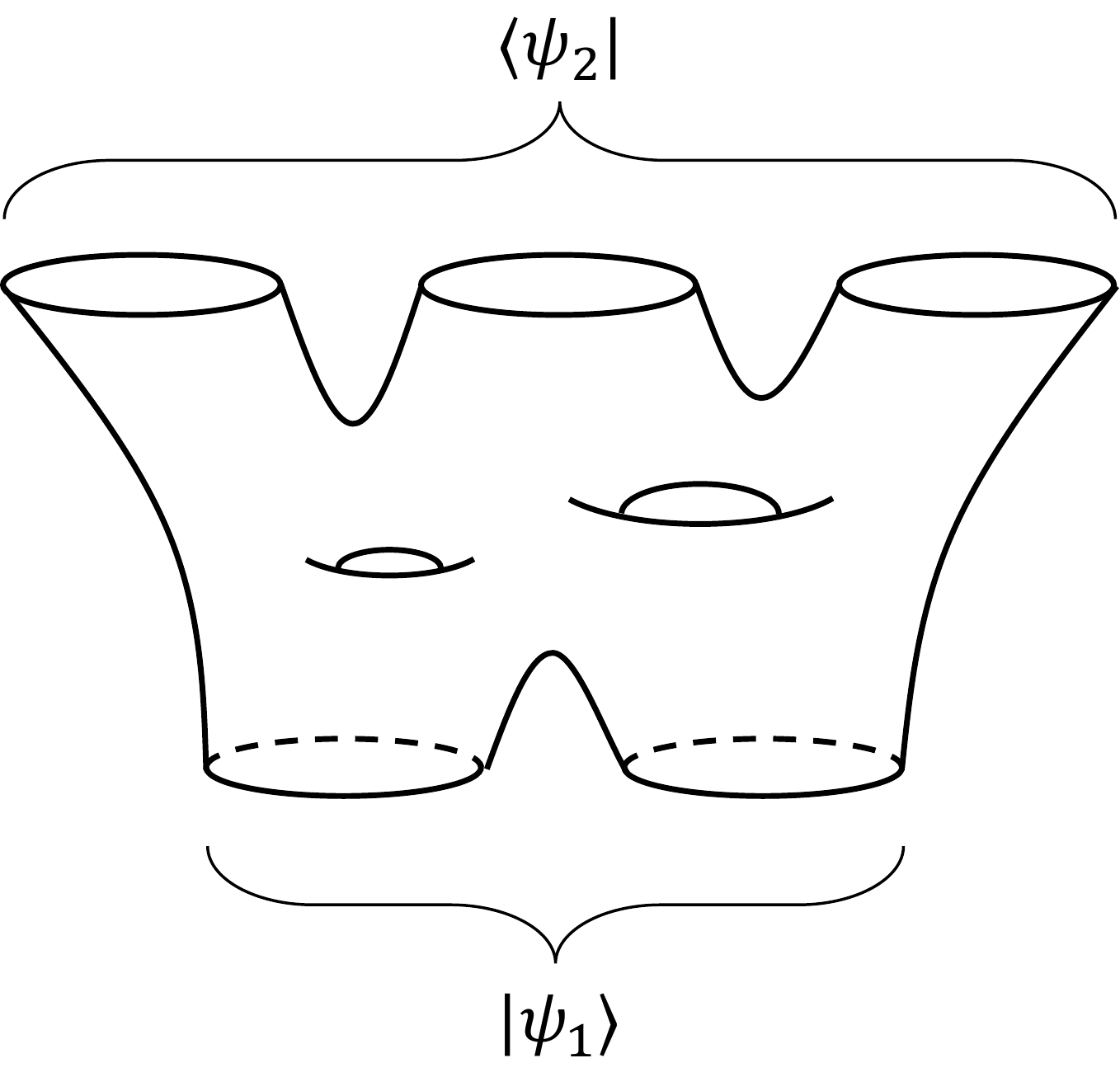}
\caption*{A contribution to the inner product $\braket{\psi_2 | \psi_1}$.}
\end{figure}

In the case of the baby universe Hilbert space, as has been recently discussed in \cite{MarolfMaxfield}, these gauge redundancies may be computed as follows, in close analogy with the reconstruction of the Hilbert space from Euclidean correlators in axiomatic quantum field theory. Heuristically, we start with a larger vector space $\widetilde \H_\BU$ of putative states of baby universes, given by finite quantum superpositions of classical states of the baby universes (in the context of AdS/CFT, \cite{MarolfMaxfield} describes these states by fixing the asymptotic boundary conditions in the Euclidean past). The Eucludean path integral over geometries and topologies defines an inner product on this vector space, which is positive semi-definite by reflection positivity. Taking the Hilbert space completion yields the Hilbert space $\H_\BU$ of baby universes.

If the inner product computed on $\widetilde \H_\BU$ is not positive definite, taking the completion includes a quotient by the space of null states, namely baby-universe wavefunctions of zero norm. Thus, wavefunctions differing by a null state yield the same physical state in $\H_\BU$, which may be viewed as a form of gauge redundancy. An important difference from the case of an ordinary gauge theory is that the physical Hilbert space cannot be viewed as arbitrary wavefunctions of some gauge-fixed variables, at least not in any obvious way. This is related to the fact that while the constraint equation in ordinary gauge theory is a first-order differential equation, whose solutions correspond to wavefunctions on the quotient space, the Wheeler-de Witt equation is second order, and induces quantum redundancies between wavefunctions of a more complicated nature \cite{Jafferis2}.

\section{Swampland Arguments for the Hypothesis}\label{swampland}

In this section, we argue for the Baby Universe Hypothesis on the basis of the Swampland Program. As a result, we conclude that in $d > 3$, the gauge redundancies described in the previous section are so great that they cut the baby universe Hilbert space down to a single state. First, we review how the condition of no free parameters should be naturally viewed as part of the statement that there are no generalized global symmetries in quantum gravity. We then argue that nontrivial $\alpha$-parameters would violate this condition. We provide further evidence that the Baby Universe Hypothesis fits naturally into the growing web of swampland conjectures, by showing that it implies the absence of ordinary global symmetries in quantum gravity. Finally, we derive a further consequence of the Baby Universe Hypothesis by applying a state-operator correspondence, concluding that there are no well defined bulk operators in a complete theory of quantum gravity.

\subsection{Free Parameters and $(-1)$-Form Symmetries}

Our experience with string theory suggests that there cannot be any free parameters in quantum gravity in dimension $d > 3$. What we mean by this is the following. Suppose we have a low energy effective theory of quantum gravity in $d >3$ with asymptotic Minkowski or AdS boundary conditions. There can be various coupling constants appearing in the effective action, which define distinct theories or superselection sectors in infinite volume. However, in order to satisfy the condition of no free parameters, these coupling constants must be identified with the asymptotic values of dynamical fields, such that the bulk interactions in the effective field theory only depend on the local values of these scalar fields. In string theory, this is realized even more strongly than might be expected, as even discrete parameters (such as the coefficients of Chern-Simons terms not fixed by supersymmetry or anomaly cancelation, or the rank of a gauge group) should be viewed as dynamical, as they tend to correspond to things like the number of branes wrapping a cycle, which of course are dynamical.

This condition should be viewed as a natural part of the statement that quantum gravity has no generalized global symmetries. In particular, a free parameter of a theory can reasonably be called a $(-1)$-form global symmetry, as has been discussed at length in the context of anomalies in the space of coupling constants \cite{CordovaFreedLamSeiberg1, CordovaFreedLamSeiberg2}. In order to see this, note that a continuous free parameter $\lambda$ in the Lagrangian is naturally associated with a $d$-form operator $\L_\lambda$, such that the effective action takes the form
\begin{equation} S(\lambda) = S_0 + \lambda \int \L_\Lambda. \end{equation}
Since $\L_\lambda$ is a top form, it is naturally closed, and so we may define a $0$-form conserved current
\begin{equation} J_\lambda = \star\ \L_\lambda, \end{equation}
which we may view as the Noether current for the associated $(-1)$-form symmetry. One objection to this interpretation is that there is no group in sight, and indeed free parameters merely vary over a smooth manifold, not a Lie group. However, this is natural, since the $d$-volume operators that implement the ``symmetry" cannot be placed back-to-back, and so there is no need for a group law. Thus, just as $p$-form symmetries for $p \geq 1$ require more data than $0$-form symmetries, namely an abelian group law, so too do $0$-form symmetries require more data than a $(-1)$-form symmetry, namely a (not necessarily abelian) group law in the first place.\footnote{This situation should be compared to the homotopy groups of a space $X$. While $\pi_k(X)$ is an abelian group for $k \geq 2$, $\pi_1(X)$ is just a group, and $\pi_0(X)$ is just a set.}

If we couple $\L_\lambda$ to a dynamical field $\phi$ rather than a free parameter $\lambda$, in that we have
\begin{equation} S = S_0 + \int \phi\ \L_\lambda, \end{equation}
this corresponds to gauging the $(-1)$-form symmetry. The fact that the different asymptotic values of $\phi$ (which may be identified with $\lambda$) define superselection sectors for $d \geq 3$ should be compared to the fact that gauge transformations which do not vanish at infinity are actually global symmetries. Thus, the swampland condition that any free parameter in quantum gravity must be the asymptotic value of a dynamical scalar can be naturally seen as the statement that any $p$-form symmetry in quantum gravity must be gauged for the case $p = -1$. In particular, this includes both continuous and discrete parameters, just as both continuous and discrete symmetries of quantum gravity must be gauged.

\subsection{Global $(-1)$-Form Symmetries from $\alpha$-Parameters}

Taking the absence of free parameters in quantum gravity for $d > 3$ as a given, we now ask what this means for $\alpha$-parameters. As described in Section \ref{coleman}, $\alpha$-parameters exactly correspond to the coupling constants of terms in the Lagrangian for our effective field theory, and so define a $(-1)$-form symmetry of our theory. The question, then, is whether this $(-1)$-form symmetry is global or gauged. In this section, we argue that there is no ambiguity: the symmetry is global, and the $\alpha$-parameters violate the condition of no free parameters. Thus, we conclude that the Baby Universe Hypothesis follows from the condition of no free parameters in quantum gravity.

Why do we claim that $\alpha$-parameters cannot simply be the asymptotic values of dynamical fields in our effective field theory? Suppose to the contrary that they were, in that the couplings $\alpha_i \L_i$ arise as the zero modes of interactions $\phi_i \L_i$ for some dynamical fields $\phi_i$. While it would take infinite energy in order for $\phi_i$ to differ from $\braket{\phi_i}$ throughout all of space, a local variation of $\phi_i$ must be allowed as a finite energy excitation of the theory by the swampland condition of triviality of cobordism \cite{McNamaraVafa}.\footnote{For continuous $\alpha$-parameters, this follows already from taking the fields to have finite mass.} Thus, within the region where $\phi_i$ differ from their asymptotic values, local observers will observe spacetime-dependent $\alpha$-parameters. If we recall that $\alpha$-parameters are simply the eigenvalues of the baby universe operators $A_i$, we immediately see a contradiction. If $A_i$ were functions of space, then the baby universes could carry nonzero energy and momentum, which is impossible. Thus, we cannot realize continuous $\alpha$-parameters as arising from dynamical fields, and so the corresponding $(-1)$-form symmetry is global. Taking the absence of global $(-1)$-form symmetries in quantum gravity for $d > 3$ as a given, we conclude that theories with $\alpha$-parameters in $d > 3$ belong to the swampland.

\begin{figure}[!ht]
\centering
\includegraphics[height = 2.2 in]{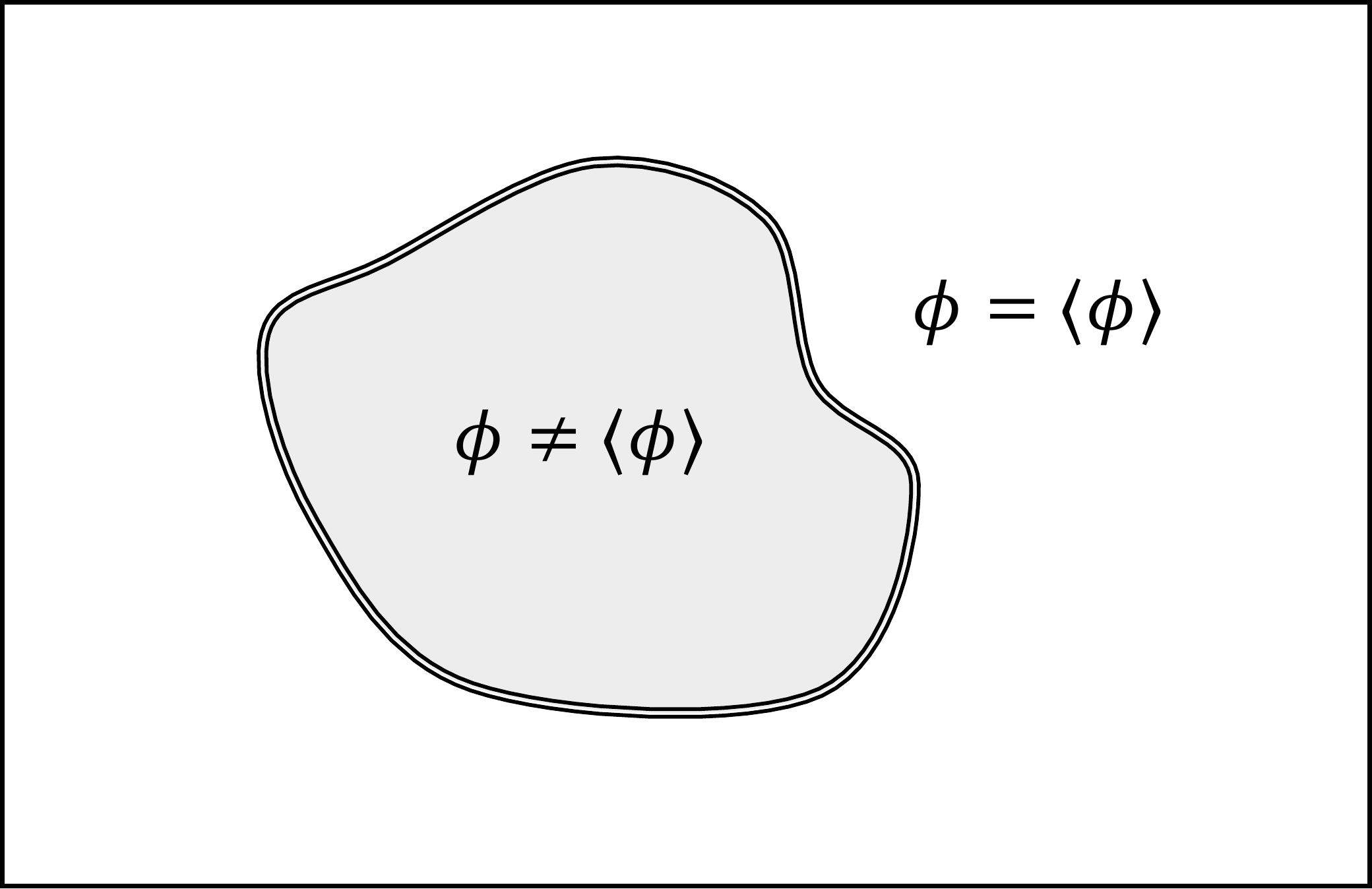}
\caption*{A finite-energy bubble where the scalar $\phi$ differs from its asymptotic value.}
\end{figure}

In fact, there is a more direct argument we could make, which is to note that $\alpha$-parameters explicitly correspond to superselection sectors in finite space. The condition of triviality of cobordism of spacetime \cite{McNamaraVafa} exactly says that this is impossible in $d > 3$, and that any superselection sectors of quantum gravity must arise as a result of distinct boundary conditions. It is important to note that while many calculations in \cite{McNamaraVafa} are done at the level of classical configurations, the statement that there cannot be superselection sectors in finite space is a statement at the level of the full quantum theory. Even if the classical theory has no such superselection sectors, it may happen that superselection sectors arise in the quantum theory, just as a quantum theory may have a symmetry that does not act on the classical configuration space. In fact, the Baby Universe Hypothesis may be viewed as a precise statement at the quantum level that there can be no superselection sectors in finite space!

\subsection{No Global Symmetries as a Consequence}

In the previous section, we have explained how previously established swampland conditions imply the Baby Universe Hypothesis. Conversely, in this section we show that the Baby Universe Hypothesis implies one of the most basic swampland conditions, that there cannot be global symmetries in quantum gravity for $d > 3$. This provides further evidence that the Baby Universe Conjecture is the natural expectation from the perspective of the Swampland Program, and indeed it fits neatly in the broader web of swampland conjectures.

The argument, in fact, is quite straightforward. Suppose for the sake of contradiction that there we had a theory of quantum gravity with a global symmetry. As Coleman notes \cite{Coleman}, while baby universes may not carry any gauge charge, they can carry charge associated with global symmetries, and so the baby universe Hilbert space must break up into the direct sum over sectors with different global charge. These different sectors are all orthogonal even after taking the gauge redundancies into account, since the Euclidean path integral cannot violate conservation of global charge by assumption (otherwise, we wouldn't have an exact symmetry). Further, these sectors are nontrivial, since we may act with a nontrivial local operator carrying the charge to move from one sector to another. Thus, the presence of a global symmetry implies that
\begin{equation} \dim \H_\BU > 1, \end{equation}
and so the Baby Universe Hypothesis implies the condition of no global symmetries.

\subsection{Operator Interpretation}

Though we have phrased the Baby Universe Hypothesis in terms of the Hilbert space of baby universe states, a further consequence follows from an application of the state-operator correspondence. Suppose we had a well-defined bulk operator supported on a compact domain in spacetime, such as a point operator or an extended operator defined on a closed manifold. Choosing a small tubular neighborhood $N$ of the operator's support, we can consider the state of quantum gravity on $\del N$ defined by a path integral with this operator insertion. By our assumption of compactness, $\del N$ is closed, and thus this state lives in the baby universe Hilbert space! Assuming the validity of the Baby Universe Hypothesis, we conclude that this state is gauge equivalent to the Hartle-Hawking wavefunction, i.e., the state produced by the insertion of the identity operator. Shrinking the neighborhood $N$ to zero size, we learn that the operator itself must be gauge equivalent to a multiple of the identity operator, and so there are no nontrivial, strictly well defined bulk operators in complete theories of quantum gravity.\footnote{This last step is actually unnecessary, since the quantum gravity path integral includes an integration over all neighborhoods $N$. Put differently, like conformal field theories or topological field theories, quantum gravity should enjoy a state-operator correspondence, as there is no background metric.} 

It is important to emphasize exactly which types of bulk operator are excluded by this conclusion. The prototypical examples of operators excluded by the Baby Universe Hypothesis are the BRST invariant vertex operators on the string worldsheet, thought of as a theory of quantum gravity in two dimensions. These operators are strictly well defined operators living at a point in spacetime, and their insertion defines a nontrivial state in the closed string Hilbert space. Identifying this closed string Hilbert space as a baby universe Hilbert space, we learn that the string worldsheet violates the Baby Universe Hypothesis, just as it violates many other swampland conditions. In general, we do not expect worldsheet theories to satisfy the Baby Universe Hypothesis, as we discuss below in Section \ref{worldsheets}.

Another important feature of the operators excluded by the Baby Universe Hypothesis is that they must be strictly well defined on the entire quantum gravity Hilbert space. In the context of AdS/CFT, there are quasi-local bulk operators in the effective field theory defined around some classical geometry. As has been argued \cite{DongHarlowWall, AlmheiriDongHarlow}, these operators are defined relative to a code subspace of the CFT, and their bulk locality only holds within that code subspace. These operators are not excluded by the Baby Universe Hypothesis, since they are not well defined local operators in the UV complete quantum theory. In fact, the arguments of \cite{HarlowOoguri1, HarlowOoguri2} can easily be extended to rule out any such strictly well defined bulk operator: such an operator would necessarily commute with all local CFT operators not on a code subspace, but on the entire CFT Hilbert space, and would thus be a multiple of the identity.\footnote{Note that this argument breaks down in the presence of an ensemble, as there would be central operators in the CFT von Neumann algebra measuring the parameters defining your ensemble.} Finally, operators whose support extends out to the AdS boundary are not excluded, since these operators do not produce a state in the baby universe Hilbert space, and need not commute with all CFT local operators.

What are we to make of the idea that there are no strictly well defined bulk observables in a complete theory of quantum gravity? At first glance, this seems like nonsense, since our day-to-day experience includes making well-defined measurements in a quantum theory of gravity. However, there is no contradiction, and the resolution is simply that quantum gravity is background independent! The bulk operators excluded by the Baby Universe Hypothesis describe probes of quantum gravity from \emph{outside} the system, that is, measurements that could be made by an observer who is not part of the universe. Instead of allowing static, background probes, quantum gravity describes a system in which every potential probe is already a dynamical object in the theory.\footnote{For example, though D-branes define static probes at zero string coupling, they become dynamical as soon as a nonzero string coupling is turned on.} The strictly well defined observables in quantum gravity do not correspond to local bulk probes, but rather to measurements made at an asymptotic boundary, such as the S-matrix in flat space, or CFT observables in AdS. In addition, there should also be relational observables describing the experience of bulk observers who are part of the quantum gravitational system. These are likely more akin to operators acting in a code subspace of the CFT, which correspond to bulk operators defined relative to an effective field theory limit.

\section{Implications for Holography}\label{holography}

 In this section, we first review how the Baby Universe Hypothesis is implied by a standard interpretation of the AdS/CFT dictionary, as well as how it resolves the potential paradoxes with factorization of correlation functions (both noted in \cite{MarolfMaxfield}). We also explain why an ensemble interpretation of holography in $d = 2$ or $3$ such as JT gravity is actually expected based on the standard (non-ensemble) version of holography in higher dimensions, and why we do not expect an ensemble to play a role for holography in $d>3$. Finally, we explain how the Baby Universe Hypothesis allows us to think of the holographic principle as a form of ``Gauss's law for entropy" in quantum gravity.

\subsection{The Standard AdS/CFT Dictionary and Boundary Locality}\label{standard_adscft}

Suppose we consider a theory of AdS quantum gravity in $(d+1)$ spacetime dimensions. If we fix asymptotic boundary conditions for our bulk theory, in the form of a conformal $d$-manifold $X$ and boundary conditions $J$ for the bulk fields, we may compute the gravitational partition function
\begin{equation} \Z_\QG(X, J) = \int_{X, J} \D g\ \D \phi\ e^{- S(g, \phi)}, \end{equation}
by integrating over bulk metrics and field configurations with the specified boundary conditions. We will supress the argument $J$ for clarity below. While there is no obvious \emph{a priori} reason for $\Z_\QG$ to depend locally on $X, J$, the miracle of the standard (non-ensemble) AdS/CFT dictionary is that the gravitational partition function is exactly equal to the partition function of a local conformal field theory defined on the boundary,
\begin{equation} \Z_\QG(X) = \Z_\CFT(X). \end{equation}

As such, $\Z_\QG$ must satisfy the axioms of local quantum field theory. In particular, given a $(d-1)$ manifold $M$, we may define a Hilbert space $\H_\QG(M)$ of bulk states which have $M$ as their asymptotic boundary. Furthermore, given two $(d-1)$-manifolds $M_1, M_2$, we have that the Hilbert space defined by the disjoint union of $M_1$ and $M_2$ must tensor factorize,
\begin{equation} \H_\QG(M_1 \sqcup M_2) = \H_\QG(M_1) \tensor \H_\QG(M_2). \end{equation}
From the bulk perspective, this is also a mystery, since the bulk Hilbert space seems to include connected geometries which do not naively factorize. However, motivated by the example of the thermofield double state and the two-sided AdS black hole \cite{Maldacena}, we have learned that connected geometries may be realized as entangled superpositions of disconnected geometries. Thus, this factorization also relies on the same gauge redundancies \cite{MarolfMaxfield, Jafferis} described above in Section \ref{gauge}.

What is the Hilbert space $\H_\BU$ of baby universes? By definition, a baby universe state is a state of closed universes, and as such has no asymptotic boundary. Put differently, we have that
\begin{equation} \H_\BU = \H_\QG(\varnothing), \end{equation}
where $\varnothing$ denotes the \emph{empty} $(d - 1)$-manifold! Now, $\varnothing$ is the unit for disjoint union, meaning that
\begin{equation} \varnothing \sqcup M = M, \end{equation}
for any $(d-1)$-manifold $M$. Together with boundary locality, this implies that we have
\begin{equation}\label{tensor_unit}
\H_\QG(M) = \H_\BU \tensor \H_\QG(M).
\end{equation}

If all the Hilbert spaces involved here were finite-dimensional, this would immediately imply
\begin{equation} \dim \H_\BU = 1, \end{equation}
simply by taking the dimension of both sides of (\ref{tensor_unit}). Since in general the Hilbert spaces $\H_\QG(M)$ will be infinite-dimensional, we need to work a bit harder. One way to argue is in the case that the dual CFT is compact, meaning that the spectrum of the Hamiltonian $H$ on $\H_\QG(M)$ is discrete for a compact manifold $M$. In this case, we can truncate the spectrum at some energy cutoff $\Lambda$, and look at the finite-dimensional Hilbert space of states with energy $E < \Lambda$. By the Wheeler-de Witt equation, all states in $\H_\BU$ have zero energy, and so we have
\begin{equation} \H_\QG(M)\Big|_{E < \Lambda} = \H_\BU \tensor \left( \H_\QG(M)\Big|_{E < \Lambda} \right). \end{equation}
Taking the dimension of both sides again implies
\begin{equation} \dim \H_\BU = 1, \end{equation}
as desired.

A better way to argue in general is to interpret (\ref{tensor_unit}) as saying that $\H_\BU$ is a unit object for the tensor product of Hilbert spaces, which implies that $\H_\BU$ is canonically isomorphic to $\CC$. In fact, that quantization on the empty manifold gives a one-dimensional Hilbert space is one of the axioms of local quantum field theory, and is completely obvious from the path integral, since there is a unique field configuration on the empty manifold. Thus, the Baby Universe Conjecture follows immediately from a standard interpretation of the holographic dictionary and the axioms of local quantum field theory.

\subsection{Factorization of Correlation Functions}

One key piece of the standard AdS/CFT dictionary that has received a great deal of recent attention \cite{Saad:2019lba} is the factorization of CFT correlation functions. What this means is that if we compute the partition function on a disjoint union $X_1 \sqcup X_2$, then the partition function factorizes,
\begin{equation} \Z(X_1 \sqcup X_2) = \Z(X_1) \Z(X_2). \end{equation}
This is another axiom of local quantum field theory. However, from the perspective of the bulk Euclidean path integral, it is again unclear how this could be the case, as there are contributions to the disconnected partition function involving connected bulk geometries, and so in order to factorization to arise from a naive bulk path integral there must be enormous cancelations between connected and disconnected contributions to the two-boundary partition function.

One way to compute this partition function is to utilize the baby universe Hilbert space, following Marolf and Maxfield \cite{MarolfMaxfield}. In particular, fixing a boundary manifold $X$ and sources, they define a state
\begin{equation} \ket{\Z(X)} \in \H_\BU, \end{equation}
by performing the Euclidean path integral with $X$ as the asymptotic boundary in the past. Alternatively, this state may be defined by acting on the Hartle-Hawking state with the operator $\widehat \Z(X)$ that inserts an additional asymptotic boundary $X$,
\begin{equation} \ket{\Z(X)} = \widehat \Z(X) \ket{\HH}. \end{equation}
Noting that $\widehat \Z(X)$ is nothing but a baby universe operator $A_i$, we see that it must be diagonal in the $\alpha$-eigenbasis with eigenvalues $\Z_\alpha(X)$, and we may compute
\begin{equation} \ket{\Z(X)} = \sum_\alpha c_\alpha \Z_\alpha(X) \ket{\alpha}, \end{equation}
where $c_\alpha = \braket{\alpha | \HH}$ is the amplitude to be in state $\ket{\alpha}$ in the Hartle-Hawking wavefunction (which we assume is taken to be normalized).

With this machinery in hand, we may immediately compute the partition function,
\begin{equation} \Z(X) = \braket{\HH | \widehat \Z(X) | \HH} = \sum_\alpha p_\alpha \Z_\alpha(X), \end{equation}
where $p_\alpha = \abs{c_\alpha}^2$ is the probability distribution on $\alpha$ in the Hartle-Hawking ensemble. Further, we must have $p_\alpha > 0$ for all $\alpha$ (see \cite{MarolfMaxfield}), and so this will always be a nontrivial ensemble if there are nontrivial $\alpha$-parameters. Just as easily, we may compute the two-boundary partition function,
\begin{equation} \Z(X_1 \sqcup X_2) = \braket{\HH | \widehat \Z(X_1) \widehat \Z(X_2) | \HH} = \sum_\alpha p_\alpha \Z_\alpha(X_1) \Z_\alpha(X_2), \end{equation}
and so we see a lack of factorization coming from a nontrivial ensemble of $\alpha$-parameters. In order to have factorization, we could work in a fixed $\alpha$-eigenstate instead of in the Hartle-Hawking state, where we would find
\begin{equation} \braket{\alpha | \widehat \Z(X_1) \widehat \Z(X_2) | \alpha} = \Z_\alpha(X_1) \Z_\alpha(X_2) = \braket{\alpha | \widehat \Z(X_1) | \alpha} \braket{\alpha | \widehat \Z(X_2) | \alpha}. \end{equation}
This is a manifestation of the fact that fixing a choice of $\alpha$ does indeed define a consistent and local boundary quantum field theory.

If the Baby Universe Hypothesis is satisfied, then the Hartle-Hawking state, as the unique state in $\H_\BU$, is in fact itself an $\alpha$-eigenstate, and so we see that we have factorization no matter what! Put differently, if the Baby Universe Hypothesis is satisfied, then we have
\begin{equation} \ket{\Z(X)} = \Z(X) \ket{\HH}, \end{equation}
which is to say that the baby universe state produced by the Euclidean path integral with asymptotic boundary $X$ is just proportional to the Hartle-Hawking state, and so factorization follows immediately from the one-dimensionality of $\H_\BU$. One consequence of this is that the Baby Universe Hypothesis requires enormous and miraculous cancelations between contributions to the Euclidean path integral with different topologies, a highly non-generic situation from the perspective of effective field theory.

\subsection{The Role of an Ensemble in AdS/CFT}\label{worldsheets}

One strong motivation to doubt the amazing cancelations described in the previous section, which are key to the validity of the Baby Universe Conjecture, is the example of JT gravity in $d = 2$. In JT gravity, the full bulk path integral may be performed without adding any new UV degrees of freedom \cite{Saad:2019lba}, and indeed, these types of cancelations do not occur. JT gravity is dual \cite{Saad:2019lba} to an ensemble average of $d = 1$ quantum systems given by a random matrix model, leading to a consistent new model for holography involving an ensemble. Further, in this framework, the disconnected partition functions do not factorize, as is expected from an exact quantization of JT gravity \cite{HarlowJafferis}. In this section we explain why this is not surprising in the $d = 2$ context, and why we should not expect it to generalize to higher dimensions (with the possible exception of $d = 3$), from yet another perspective.

Consider a collection of $k$ quantum systems in $d = 1$, defined by $N \times N$ Hamiltonians
$H_i(A)$ for $1 \leq i \leq k$ that all depend on some background parameters $A$, leading to $k$ unitary time evolution operators
\begin{equation} U_i(A, \tau_i)= e^{-i \tau_i H_i(A)}, \end{equation}
We may then consider the ensemble average over $A$ of their partition function on $k$ copies of $S^1$ with lengths $\tau_i$, given by
\begin{equation} \Z = \int dA\ e^{-S(A)} \prod_{i=1}^k {\rm Tr}\ U_i(A, \tau_i) \end{equation}

We have not yet stated what $A$, $H_i(A)$, and $S(A)$ are; we now fill this gap. Consider a $D$-dimensional sphere with $k$ circles $\gamma_i$ of lengths $\tau_i$ on it, and consider an $SU(N)$ gauge connection $A$ on $S^D$. We identify 
\begin{equation} U_i(A, \tau_i) ={\rm P}\ {\rm exp}\left({i \int_{\gamma_i}A}\right), \end{equation}
as the Wilson line observables in the gauge theory. Moreover, we identify
\begin{equation} S(A) = \frac{1}{g^2} \int {\rm Tr} (F \wedge \star F).\end{equation}
We can now reinterpret $\Z$ as the expectation value of $k$ Wilson loop observables in a $D$-dimensional $SU(N)$ gauge theory,
\begin{equation} \Z = \left\langle \prod_i {\rm Tr} \ U_{\gamma_i } \right\rangle. \end{equation}
In other words, from the perspective of the $D$-dimensional theory, the $d = 1$ ensemble average is nothing but the path integral description of the $SU(N)$ gauge theory.

Now suppose this theory has a large $N$ dual (if necessary we can consider adding additional fields and couplings to the above scenario to realize this). Then we can ask how to describe the large $N$ dual theory in terms of sum over Riemann surfaces with $k$ boundaries, as suggested by `t Hooft \cite{tHooft2}. In the context of standard (non-ensemble) version of AdS/CFT in string theory, the corresponding `t Hooft surfaces are identified with string worldsheets, which can be viewed as a $d = 2$ quantum gravitational system. Summing over all Riemann surfaces with $k$ boundaries is the standard large $N$ description of $\Z$, so we learn that {\it the Wilson loops observables of any large $N$ gauge theory which admits a dual description leads to a $d = 2$ theory of quantum gravity which computes the average of these $d = 1$ partition functions over an ensemble!}\footnote{See \cite{BetziosPapadoulaki} for a similar perspective in the context of the $c = 1$ string.} So we seem to have reproduced a scenario analogous to JT gravity in a much more general context. We now explain more explicitly how JT gravity fits in this picture, which was the motivation for this general construction.

Consider $SU(N)$ Chern-Simons gauge theory on $S^3$. It is known that this theory can be realized by topological A-model strings on $T^*S^3$ with $N$ topological branes wrapping $S^3$ \cite{Witten:1992fb}. Using this setup, it has been argued in \cite{Gopakumar:1998ki} that this has a large $N$-dual given by A-model topological gravity on the resolved conifold. In this context, the worldsheet diagrams of topological strings can be viewed as large $N$ `t Hooft surfaces. In computing $k$ Wilson loop observables, we would be considering diagrams with $k$ boundaries (for examples of this see \cite{Ooguri:1999bv}, as in the general scenario above). Moreover, the mirror of this picture leads to B-model topological string description. In particular, one finds that the large $N$ limit of this matrix model is holographically dual to topological gravity on the resolved conifold given by the Calabi-Yau 3-fold:
\begin{equation}x^2+y^2+u^2+v^2=0.\end{equation}
As it was explained in \cite{Dijkgraaf:2002fc} this can be generalized to an arbitrary matrix model whose gravity dual is given by a local Calabi-Yau 3-fold
\begin{equation} F(x,y)+u^2+v^2=0, \end{equation}
where $F(x,y)=0$ describes the spectral curve of the matrix model. JT gravity is related \cite{Saad:2019lba} to the partition function of Mirzakhani's model \cite{Mirzakhani} which in turn is given by the matrix model with spectral curve \cite{Eynard:2007fi},
\begin{equation} F(x,y)=y^2-\sin^2{\sqrt x}. \end{equation}
For an explanation of this based on the relation of Mirzakhani's model to Mumford classes see \cite{Dijkgraaf:2018vnm}.\footnote{ The connection of this theory with Mirzakhani's model can be explained by noting that the $(1,2)$ minimal model coupled to $d = 2$ topological gravity leads to computation of Mumford classes \cite{Witten:1989ig}, and that this theory before deformation is desribed by the topological B-model on $y^2-x+u^2+v^2=0$ \cite{Aganagic:2003qj}.  
Including deformations to convert Mumford classes to Mirzakhani model is equivalent to replacing $x$ with $\sin^2{\sqrt x}$ as explained in \cite{Dijkgraaf:2018vnm}.  It is interesting that Mirzakhani's model can be viewed as computing the partition function of a topological string on a non-compact Calabi-Yau 3-fold.}  Therefore we see that
JT gravity is equivalent to the worldsheet description of topological gravity on the non-compact Calabi-Yau 3-fold given by
\begin{equation} y^2-\sin^2{\sqrt x }+u^2+v^2=0. \end{equation}
Moreover it is explained in \cite{Dijkgraaf:2007sx} 
how the Eynard-Orantin rules of computation for large $N$ dual of matrix models \cite{Eynard:2007kz} can be interpreted as arising from the $d = 6$ topological gravity of the B-model, which is the Kodaira-Spencer theory of gravity in 6 dimensions \cite{Bershadsky:1993cx}. Another example of this type is the model studied in \cite{BetziosPapadoulaki} of the $c = 1$ string, which is equivalent at the self-dual radius to the B-model on the deformed conifold \cite{GhoshalVafa}.

Thus, we have explained why one can think of the average over an ensemble of background parameters for a system of $d = 1$ as naturally induced from computation of specific correlators of a higher dimensional gauge theory, and that the emergence of the $d = 2$ quantum gravity should be viewed as a large N dual description ala `t Hooft. From this perspective, we can ask whether there can be a higher dimensional generalization to $d \geq 3$ where we get an ensemble average? To have a higher dimensional generalization along these lines, we would need to replace the `t Hooft surfaces with higher dimensional objects whose boundaries are related to observables of the corresponding quantum system. However, the only known quantum systems involve either particles or strings, leading to observables involving loops or surfaces.
So at most we may expect to generalize this $d = 2$ story to $d = 3$, where, say, M2 branes may be viewed as replacing the role of string worldsheets for the theory of large $N$ M5 branes, ending on the surface operators of this theory. Of course there is currently no known way of thinking of M2 branes in the context of M-theory holography as playing the same role that strings play for the large $N$ description of gauge theories. Regardless of whether this can be viewed as a $d = 3$ gravitational system, this perspective clearly suggests that this example cannot be generalized to $d>3$, reinforcing the arguments we presented based on other swampland principles. It is quite interesting that in this context the fact that $d=2$ and perhaps $d=3$ may be exceptional cases for swampland principles is mirrored by the fact that the only known quantum systems decoupled from gravity involve either particles or strings.

\subsection{Holography as Gauss's Law for Entropy}\label{gauss_law}

While many discussions of holography occur within the context of AdS/CFT, there is a more basic holographic principle, due to `t Hooft \cite{tHooft} and Susskind \cite{Susskind}, that should apply in much more generality. In this section, we explain how the Baby Universe Hypothesis naturally leads us to a picture where we interpret the holographic principle as providing a form of ``Gauss's law for entropy" in quantum gravity.

What does Gauss's law tell us? At its most basic, say in the context of electromagnetism, Gauss's law comes in two versions, a local form and a global form. The local form tells us that we may measure the charge inside any region by a local calculation on the boundary of the region, namely by computing the flux of the electric field through the boundary. The global form tells us that if we take space to be a closed manifold, the net charge must vanish, since electric field lines emanating from a positive charge must end somewhere on an equal and opposite negative charge. Of course, the local and global forms are connected: since the boundary of a closed space is empty, there is no way for it to support a nonzero electric flux.

The analogy to the holographic principle is straightforward. In particular, the holographic principle tells us that the entropy of a bulk region of space may be computed as the entropy of a state in a local quantum system living on the boundary of the region. This is a local statement, and should be compared to computing the charge inside a region by the flux of the electric field through the boundary. What, then, is the global analog? We claim that it is precisely the Baby Universe Hypothesis! Indeed, saying that the Hilbert space of baby universes is one-dimensional means that any quantum state of closed universes carries no entropy. The arguments in Section \ref{standard_adscft} that identify the baby universe Hilbert space with the Hilbert space of the dual quantum field theory on the empty manifold are analogous to the derivation of the global version of Gauss's law from the local version.

It is important to note the key role played by quantum mechanics in this analogy. In particular, unlike for classical systems, quantum entanglement allows the entropy of a collection of quantum subsystems to be smaller than the sum of the entropies of each subsystem. In the context of quantum gravity, this may be realized as follows. If we imagine cutting up a closed universe into many regions with boundaries, each boundary can have nonzero area, and so each region can have nonzero entropy. However, in order to glue the subregions together into a closed universe, we must choose specific entangled states on the boundaries (corresponding to sewing the geometries together), and in this way end up with a much lower entropy (namely, zero) than the sum of the entropies of each region, as required by the Baby Universe Hypothesis.

\section{Conclusion}\label{conclusion}

In this note, we argued that the Baby Universe Hypothesis, which was noted in \cite{MarolfMaxfield} as one potential resolution to the paradoxes of the Euclidean path integral, is in fact the natural resolution from the perspective of the Swampland Program. Further, it provides a clean synthesis of many things we hope to be true about quantum gravity, including the validity of the Euclidean path integral, the absence of free parameters in $d > 3$, and the standard understanding of the AdS/CFT dictionary. We interpreted the possibility of an ensemble average in $d = 2$ and potentially $3$ as arising naturally from worldvolume perturbative expansions of a larger theory of quantum gravity that does indeed satisfy the Baby Universe Hypothesis. Thus, while we cannot prove the truth of the Baby Universe Hypothesis (since we do not have a complete theory of quantum gravity with $d >3$), it is our belief that it indeed holds in higher-dimensional and complete theories of quantum gravity, and in our universe in particular.

\section*{Acknowledgments}

We would like to thank Robbert Dijkgraaf, Dan Freed, Arthur Hebecker, Simeon Hellerman, Daniel Jafferis, Juan Maldacena, Donald Marolf, Henry Maxfield, Miguel Montero, Georges Obeid, Steve Shenker, Pablo Soler, and Irene Valenzuela for useful discussions.
We have greatly benefited from the hospitality of UC Santa Barbara KITP where this project was completed.

 The research of C.V. is supported in part
by the NSF grant PHY-1719924 and by a grant from the
Simons Foundation (602883, CV).  This research was supported in part by the National Science Foundation under Grant No. NSF PHY-1748958.
This material is based upon work supported by the National Science Foundation Graduate
Research Fellowship Program under Grant No. DGE1745303. Any opinions,
findings, and conclusions or recommendations expressed in this material are those of the
authors and do not necessarily reflect the views of the National Science Foundation.

\end{document}